\documentclass[a4paper,11pt]{article}

\usepackage{amsmath, amssymb, bm}
\usepackage[]{graphicx}
\usepackage[font=scriptsize,labelfont=bf]{caption}
\usepackage[caption=false]{subfig}
\usepackage{slashed}
\usepackage[top=1in, bottom=1.2in, left=0.7in, right=0.7in]{geometry}
\usepackage{authblk}
\usepackage{hyperref}

\newcommand{\bs}{\boldsymbol}

\begin{document}

\title{\bf{Axion mediated photon to dark photon mixing}}
\author{Damian Ejlli}

\affil{\emph{\normalsize{Department of Physics, Novosibirsk State University, Novosibirsk 630090 Russia and \\ \emph{Theory Group, Laboratori Nazionali del Gran Sasso, 67100 Assergi, L'Aquila Italy}}}}

\date{}

\maketitle

\begin{abstract}
The interaction between the dark/mirror sector and the ordinary sector is considered, where the two sectors interact with each other by sharing the same QCD axion field. This feature makes the mixing between ordinary and dark/mirror photons in ordinary and dark electromagnetic fields possible. Perturbative solutions of the equations of motion describing the evolution of fields in ordinary and dark external magnetic fields are found. User-friendly quantities such as transition probability rates and Stokes parameters are derived. Possible astrophysical and cosmological applications of this model are suggested.

\end{abstract}


\vspace{1cm}

\section{Introduction}
\label{sec:1}

One main problem in quantum chromodynamics (QCD) is that it preserves the charge-parity (CP) symmetry which is observed to be broken in weak interactions. In general, if we are not concerned with the violation of the CP symmetry or time (T) symmetry, in any gauge theory we can introduce in the Lagrangian density a term of the type $\mathcal L\propto \theta_{\alpha\beta}\,\epsilon^{\mu\nu\rho\sigma}F_{\mu\nu}^\alpha F_{\rho\sigma}^\beta$, where $\theta_{\alpha\beta}$ is a constant matrix and $F_{\mu\nu}^{\alpha}$ is a gauge field tensor. In the case of QCD, the P, T, and CP violating term in the Lagrangian density is $\mathcal L_\theta\propto \bar\theta\, G_{\mu\nu, a}\tilde G^{a, \mu\nu}$ where $\bar\theta$ is the effective angle of the theory and $G_{a, \mu\nu}$ is the gluon field tensor. However, one problem is that the CP violating term induces electric dipole moments in baryons, where for example in the case of the neutron, theoretical estimates give for the dipole moment $d_n(\bar\theta)\simeq 10^{-16}\bar\theta\, e$ cm \cite{Baluni:1978rf} while experimentally is found $d_n<2.9\times 10^{-26} e$ cm \cite{Baker:2006ts}. Such a small experimental value for $d_n$ implies a small  effective angle of the order $\bar\theta \lesssim10^{-10}$, namely the so called strong CP problem.

One possible solution for the strong CP problem, is based on the Peccei-Quinn (PQ) mechanism \cite{Peccei:1977hh} where the existence of a new particle, the axion, is proposed. In the PQ mechanism, $\bar\theta$ becomes a dynamical field with an effective potential $V(a)$ for the axion field $a$ induced by non-perturbative QCD effects. The vacuum expectation value (VEV) of the axion field $\langle a\rangle =-\bar\theta f$ is minimum for the effective potential and the CP violating term in the effective Lagrangian is dynamically cancelled. Usually, the axion scale parameter $f$ is a free one and is model dependent. Originally, $f$ was taken to coincide with the electroweak scale \cite{WW} but the non observation of axions in experiments would suggest that its scale could in principle be much larger than the electroweak scale. This fact has been implemented in the so called invisible axion models, namely the KSVZ axion model \cite{Shifman:1979if} and DFSZ axion model \cite{Dine:1981za}.

A more complicated possibility that  solves the strong CP problem is based on the mirror symmetry, namely M-symmetry\footnote{The first proposed solution for the strong CP problem based on M-symmetry, was considered in Ref. \cite{Rubakov:1997vp} in the context of complex grand unification theories, namely non supersymmetric GUT based on the gauge group $SU(5)\times SU^\prime(5)$.} see Ref. \cite{Berezhiani:2000gh}. This possible solution of the strong CP problem is still based on the PQ mechanism but the particle content group is duplicated with respect to the standard model (SM), namely one adds an additional sector of particles, the mirror sector. In this context the strong CP problem is solved simultaneously in both sectors through the PQ mechanism where the two sectors are supposed to weakly interact with each other, primarily through the gravitational force.

The general idea of the M-symmetry is based on the assumption that there exist a parallel sector of mirror or dark particles which has the same group and coupling constants analogous to the SM sector \cite{Foot:1991bp}. In this model the SM Lagrangian is invariant under M-symmetry. More precisely,  the gauge group of the theory is $G\times G^\prime$ where $G$ is the ordinary group of the SM of particles $G=SU(3)\times SU(2)\times U(1)$ with fermion fields $\Psi_i= q_i, l_i, \bar u_i, \bar d_i, \bar e_i$ and Higgs doublets $H_1, H_2$ and $G^\prime=SU(3)^\prime\times SU(2)^\prime\times U(1)^\prime$ is the mirror gauge group\footnote{In this paper the sign $(^\prime)$ denotes quantities of the mirror sector if not otherwise specified. } with analogous particle content $\Psi_i^\prime= q_i^\prime, l_i^\prime, \bar u_i^{\prime}, \bar d_i^{\prime}, \bar e_i^{\prime}$ and Higgs doublets $H_1^\prime, H_2^\prime$. Here, $q_i, i=1, 2, 3$ is the left handed quark doublet, $l_i$ is the left handed lepton doublet, $\bar u_i$ is the right handed quark singlet ($u, c, t$), $\bar d_i$ is the right handed quark singlet ($d, s, b$) and $\bar e_i$ is the right handed anti-lepton singlet. Here fermions are represented as Weyl spinors.

In the case when M-parity is an exact symmetry, the particle physics must be the same in both sectors. For example, for the Yukawa theory, we would see that ordinary and mirror sectors have the same pattern $\mathcal L_\textrm{Yuk}=Y_{U}^{ij}\bar u_iq_jH_2+Y_D^{ij}\bar d_iq_jH_1+Y_E^{ij}\bar e_il_jH_1+h.c.,\, \mathcal L_\textrm{Yuk}^\prime=Y_{U}^{\prime ij}\bar u_i^{\prime}q_j^\prime H_2^\prime+Y_D^{\prime ij}\bar d_i^{\prime}q_j^\prime H_1^\prime+Y_E^{\prime ij}\bar e_i^{\prime}l_j^\prime H_1^\prime+h.c.,\nonumber$
where $Y_l^{ij}=Y_l^{\prime ij}$ with $l=\{U, D, E\}$ are the Yukawa couplings ($3\times 3$ complex matrices) and are equal in both sectors. Since the Yukawa couplings are the same, this would imply that quark and lepton mass matrices have the same form, namely $\mathcal M_U=G_U\langle H_2\rangle$, $\mathcal M_U^\prime=G_U\langle H_2^\prime\rangle$, $\mathcal M_D=G_D\langle H_1\rangle$, $\mathcal M_D^\prime=G_D\langle H_1^\prime\rangle$ etc. On the other hand, the total renormalizable Higgs potential in this model has the form $ \mathcal V_\text{tot}=\mathcal V+\mathcal V^\prime+\mathcal V_\text{mix}$,
where $\mathcal V$ is the standard model Higgs potential and $\mathcal V^\prime$ is the mirror/dark sector Higgs potential with the same pattern as its standard model counterpart. The mixing potential comes out due to gauge symmetry of the theory and has a quartic interaction term of the form
$\mathcal V_\text{mix}=-\kappa (H_1H_2) (H_1^\prime H_2^\prime)^\dagger+h.c.,$ where the coupling constant $\kappa$ is real due to M-symmetry. 

The M-parity can be spontaneously broken with the introduction of a real scalar singlet $\eta$ with odd parity, namely under the M-parity it changes the sign $\eta\rightarrow -\eta$. If $\eta$ has a non-zero VEV, namely $\langle\eta\rangle=\mu$, it would induce differences in mass-squared of ordinary and mirror Higgses. This difference would imply that VEVs, $v_{1, 2}$ are different from $v_{1, 2}^\prime$ and consequently we would have different weak interaction scales $v\neq v^\prime$ where $v=(v_1^2+v_2^2)^{1/2}\simeq 247$ GeV and $v^\prime=(v_1^{\prime 2}+v_2^{\prime 2})^{1/2}$.

In the M-symmetry solution of the strong CP problem, the axion field is identified as a linear combination of the Higgs doublets phases $\phi$ and $\phi^\prime$ with $a=f_a^{-1}(f\phi+f^\prime\phi^\prime)$ where $f_a$ gets contribution from both ordinary and dark sectors, $f_a=\sqrt{f^2+f^{\prime 2}}$, with $f^\prime=v_1^\prime v_2^\prime/v^\prime$ being the axion decay constant in the dark sector and $f=v_1 v_2/v$ being the axion decay constant in the ordinary sector; see Ref. \cite{Berezhiani:2000gh} for details. Consequently, the axion mass $m_a$ gets contribution from ordinary and dark sectors 
\begin{equation}\label{ax-mass}
m_a^2=\frac{N^2}{f_a^2}\left(\frac{VK}{V+K\,\text{Tr} \mathcal M^{-1}}+\frac{V^\prime K^\prime}{V^\prime+K^\prime\,\text{Tr} \mathcal M^{\prime-1}}\right),
\end{equation}
where $N$ is the color anomaly of $U(1)_\text{PQ}$ current, $K$ and $K^\prime$ are, respectively, the gluon condensates of ordinary and dark sectors which are, respectively, related to the ordinary and dark QCD scales $\Lambda, \Lambda^\prime$ through $K\sim \Lambda^3, K^\prime\sim \Lambda^{\prime 3}$ and $V, V^\prime$ are, respectively, the quark condensates of ordinary and dark sectors with $V\sim \Lambda^3$ and $V^\prime\sim \Lambda^{\prime 3}$. Here $\mathcal M$ and $\mathcal M^\prime$ are, respectively, the mass matrices of light quarks of ordinary and dark sectors where $\mathcal M=\text{diag} (m_u, m_d)$ and $\mathcal M^\prime=\text{diag}(m_u^\prime, m_d^\prime)$.

One characteristic of this model is that in the case when $f^\prime\gg f$ or $\Lambda^\prime\gg \Lambda$, the axion field $a$ couples to ordinary sector as DFSZ-like axion while it couples to the dark sector as the original axion or Weinberg-Wilczek (WW) axion \cite{WW}. In this case while the axion behaves as DFSZ-like axion with respect to the ordinary sector its mass given in \eqref{ax-mass} gets contribution from a small term coming from the ordinary sector and a much larger term coming from the dark sector. In addition, the axion field couples to photons\footnote{In this work we call the photon of the mirror sector simply dark photon. In the literature also the name hidden photon for the dark/mirror photon is used.} with two different coupling constants $g_{a\gamma}$ and $g_{a\gamma}^\prime$, which are, respectively, given by 
\begin{equation}\label{coup-cons}
g_{a\gamma}\simeq \frac{\alpha_S}{\pi}\frac{Nz}{f_a(1+z)}, \quad g_{a\gamma}^\prime\simeq \frac{\alpha_S}{\pi}\frac{Nz^\prime}{f_a(1+z^\prime)},
\end{equation}
where $z=m_u/m_d$, $z^\prime=m_u^\prime/m_d^\prime$ and $\alpha_S$ is the fine structure constant.

In itself, the introduction of the mirror sector can have several consequences in cosmology \cite{Blinnikov:1982eh} and consequently there exist several constraints on the main parameters of the model, and for a detailed review see Ref. \cite{Berezhiani:2003xm}. The application of the M-symmetry does not necessarily means that the abundances of mirror sector particles are the same as those of the ordinary sector. On the contrary, the abundances of elements of ordinary and mirror sectors must be different, not necessarily for all elements, in order to avoid any conflict with well known constraints on extra degrees of freedom such as those imposed by big bang nucleosynthesis (BBN) etc. Indeed, the BBN constraint on the number of extra degrees of freedom, which usually is expressed in terms of the effective neutrino species, constraints the mirror sector equilibrium temperature $T^\prime$ to be $T^\prime<0.64\, \Delta N_\nu^{1/4} T$ where $\Delta N_\nu$ is the effective number of neutrino species and $T$ is the equilibrium temperature of the ordinary sector. The fact that $T^\prime<T$, means that the mirror and ordinary sectors do not come in thermal equilibrium and therefore they evolve almost separately, a condition which is easily achieved if the two sectors communicate through the gravity force. Another constraint imposed on the parameters of the model comes from the mixing term $\mathcal V_\text{mix}$ of the ordinary and mirror sector Higgs doublets. The presence of such term in the Lagrangian density, would make possible the decay $H_{1, 2}^\dagger H_{1, 2}\rightarrow H_{1, 2}^{\prime\dagger} H_{1, 2}^\prime$, which in principle would bring the two sectors in equilibrium in the early universe unless $\kappa$ is very small, namely $\kappa<10^{-8}$ \cite{Berezhiani:1995am}.

Another important consequence with the introduction of the mirror sector is that it may provide the right abundance of elements in order to explain the origin of dark matter in a rather natural way. Indeed, as shown in Refs. \cite{Berezhiani:2000gw}-\cite{Bento:2001rc}, it is possible that the baryon asymmetry in the early universe for the mirror sector could be larger than that of the ordinary sector and consequently the number density of mirror baryons would be larger than that of the ordinary sector, namely $n_B^\prime\geq n_B$. In the case when $n_B^\prime/n_B\simeq 5$, we would see that the mirror particles would be plausible candidates for the dark matter; see Ref. \cite{Berezhiani:2003xm} for details.

The solution of the strong CP problem through the PQ mechanism in both sectors and the introduction of the axion field which communicates simultaneously with the ordinary and dark sectors, give a unique possibility to explore the vast implications of the model. As I will show in this work, an important consequence of the model proposed in Ref. \cite{Berezhiani:2000gh} is that ordinary photons can mix with dark photons by sharing the same axion field. Such process is very important especially in those situations where do exist both ordinary and dark external magnetic fields. In this case is possible for dark photons to transform into ordinary photons and vice-versa in external magnetic fields. Such mixing/oscillation is very important in the early universe where in the presence of ordinary and dark large-scale magnetic fields, the dark CMB photons would mix/oscillate into ordinary CMB photons and vice versa. This situation could in principle be realized in the early universe since there are enough left ordinary and mirror baryons that can contribute to the generation of large-scale magnetic fields. In addition, the photon-axion-dark photon mixing/oscillation would be important also in those situations where dark objects emit dark photons into intergalactic space where both ordinary and dark large-scale magnetic fields might coexist.

In this work, I present a model in which the two sectors interact only via the same axion field in the case when ordinary and dark external magnetic fields coexist in the same place and at the same time. Here I assume the large-scale dark magnetic field to be generated in an analogous way as the ordinary large-scale magnetic field. In addition, I consider the axion mass given in expression \eqref{ax-mass} to be a free parameter of the model without any a priory assumption if the biggest contribution to $m_a$ comes either from the ordinary sector or from the dark sector. This work is organized as follows: in Sect. \ref{sec:2}, I introduce the model of photon-axion-dark photon mixing and derive the field equations of motion in external magnetic fields. In Sect. \ref{sec:3}, I calculate the transitions probabilities for different transition channels and calculate the Stokes parameters which describe the polarization state of the light. In Sect. \ref{sec:4}, I suggest some possible applications of the proposed model and in Sect. \ref{sec:5}, I conclude. In this work I adopt the metric with signature $\eta_{\mu\nu}=\text{diag}({1, -1, -1, -1})$ and work with the natural (rationalized) Lorentz-Heaviside units ($k_B=\hbar=c=\varepsilon_0=\mu_0=1$) with $e^2=4\pi \alpha$.

\section{Photon-axion-dark photon mixing: the model}
\label{sec:2}

\begin{figure}[htbp]
\begin{center}
\includegraphics[scale=0.55]{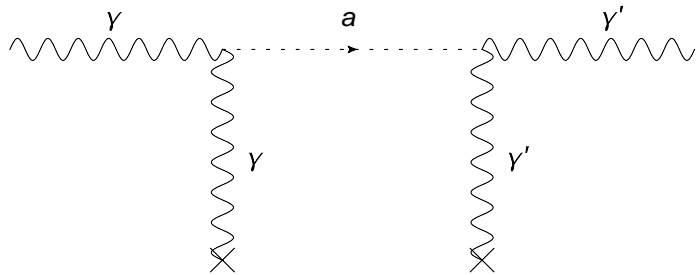}
\caption{Axion mediated photon to dark photon transition in ordinary and dark external magnetic fields. The external magnetic fields are denoted with cross symbols.}
\label{fig:1}
\end{center}
\end{figure}

The M-symmetry solution of the strong CP problem introduced in Sect. \ref{sec:1}, has several interesting theoretical and phenomenological aspects. Before proceeding further, is necessary to stress right now that apart from interacting with the same axion field $a$, the two sectors also interact gravitationally but this interaction is not important for the purposes of this work and will not be considered in what follows. In particular, in this work we are mostly interested in the interaction of the axion field with ordinary and dark photon fields. Therefore, let us consider the model where the effective Lagrangian density is given by
\begin{equation}\label{tot-act}
\mathcal L_\text{eff}=-\frac{1}{4}F_{\mu\nu}F^{\mu\nu}+\frac{1}{4}g_{a\gamma}\,a\,F_{\mu\nu}\tilde F^{\mu\nu}-\frac{1}{2}m_a^2\,a^2+\frac{1}{2}\partial_\mu a\partial^\mu a -\frac{1}{4}F_{\mu\nu}^\prime F^{\prime\mu\nu}+\frac{1}{4}g_{a\gamma}^\prime\,a\,F_{\mu\nu}^\prime\tilde F^{\prime\mu\nu}+\mathcal L_\text{med}
\end{equation}
where $F_{\mu\nu}$ is the electromagnetic field tensor of the ordinary sector and $F_{\mu\nu}^\prime$ is the electromagnetic field tensor of the dark sector. We may note the appearance of the axion field $a$ in the second and sixth terms in expression \eqref{tot-act} which make possible the mixing of ordinary photons with dark photons mediated by $a$; see Fig. \ref{fig:1}. The last term in \eqref{tot-act} is the interaction Lagrangian of photons and dark photons with ordinary and dark media. Such term essentially corresponds to the forward scattering of photons and dark photons in media which is encoded in the index of refraction. Generally, the Lagrangian density in this case involves a non local photon and dark photon polarization tensors in position space and is given by \cite{Melrose}
\begin{equation*}
\mathcal L_\text{med}=-(1/2) \int d^4 x^\prime A_\mu(x) \Pi^{\mu\nu}(x, x^\prime) A_\nu(x^\prime)-(1/2) \int d^4 x^\prime A_\mu^\prime(x) \Pi^{\prime \mu\nu}(x, x^\prime) A_\nu^\prime(x^\prime),\end{equation*}
where $A_\mu, A_\mu^\prime$ are, respectively, the ordinary and dark photon fields and $\Pi^{\mu\nu}, \Pi^{\prime \mu\nu}$ are, respectively, the photon polarization tensors of ordinary and dark photons in ordinary and dark media. The interaction Lagrangian $\mathcal L_\text{med}$ gives rise to dispersion relations for the ordinary and dark photons in ordinary and dark media. 

The equations of motions of \eqref{tot-act} for the fields $A^\nu, A^{\prime \nu}$ and $a$ in the case when particles propagate in ordinary and dark media are, respectively, given by
\begin{eqnarray}\label{gen-eq-mot}
\Box A^\mu-\int d^4x^\prime\, \Pi^{\mu\nu}(x, x^\prime)\, A_\nu(x^\prime)&= &g_{a\gamma}\tilde F^{\nu\mu}\partial_\nu a,\nonumber \\
\Box A^{\prime\mu} -\int d^4x^\prime\, \Pi^{\prime\mu\nu}(x, x^\prime)\, A_\nu^\prime(x^\prime) &= &g_{a\gamma}^\prime\tilde F^{\prime\nu\mu}\partial_\nu a,\\
(\Box +m_a^2)a &=&\frac{1}{4}g_{a\gamma}F_{\mu\nu} \tilde F^{\mu\nu} +\frac{1}{4}g_{a\gamma}^\prime F_{\mu\nu}^\prime \tilde F^{\prime\mu\nu}. \nonumber
\end{eqnarray}
Next, we assume that media is magnetized, namely there is respectively an external magnetic field in ordinary and dark sectors where photons and dark photons propagate through. Adopting the Coulomb gauge \footnote{In the Coulomb gauge there are also the equations of motion for $A^0$ and $A^{\prime 0}$ (scalar potentials) which are, respectively, proportional to $(\nabla\cdot a)\bs B_e$ and $(\nabla\cdot a)\bs B_e^\prime$ for a globally neutral medium. In the case when $\bs k\cdot \bs B_e=0$ and $\bs k\cdot \bs B_e^\prime=0$ and/or $a$ is spatially homogeneous, there is not mixing of the modes $A^0$ and $A^{\prime 0}$ with the usual ordinary and dark photons transverse modes and the axion field. However, in the opposite case there is mixing of $A^0$ and $A^{\prime 0}$ with the transverse ordinary and dark photons states and the axion field, but the effects of these equations to the mixing problem are very small and can be safely neglected for our purposes \cite{Das:2004ee}. }, the equations of motions for the vector potentials $\bs A^i, \bs A^{\prime i}$ and axion field $a$ become 
\begin{eqnarray}\label{eq-A-ph}
(\partial_t^2-\nabla^2)\bs A^i+\int d^4 x^\prime\,\Pi^{ij}(x, x^\prime) \bs A_j(x^\prime) &=&-g_{a\gamma}(\partial_t a) \bs B_e^i,\nonumber\\
(\partial_t^2-\nabla^2)\bs A^{\prime i}+\int d^4 x^\prime\,\Pi^{\prime ij}(x, x^\prime) \bs A_j^\prime(x^\prime) &=&-g_{a\gamma}^\prime(\partial_t a) \bs B_e^{\prime i},\nonumber\\
(\partial_t^2-\nabla^2+m_a^2)a &=& g_{a\gamma}\partial_t\bs A_i\cdot\bs B_e^i+g_{a\gamma}^\prime\partial_t\bs A_i^\prime\cdot\bs B_e^{\prime i}.
\end{eqnarray}
In obtaining the first and second equations in \eqref{eq-A-ph}, on the left hand side we neglect the terms $\Pi^{i0}A_0$ and $\Pi^{\prime i0}A_0^\prime$ under the integral sign, which are expected to be very small and give negligible contributions to the equations of motion, see Appendix \ref{appendix-B} for a general discussion.

Let us expand the fields $\bs A^i,\bs  A^{\prime i}$ and $a$ in Fourier modes for fixed wave-vector $\bs k$ as
\begin{equation}\label{field-exp}
\bs A^i({\bs x}, t)=\sum_\lambda \bs e_i^{\lambda} A_{\lambda}({\bs k, t})e^{i \bs k \bs x}, \quad \bs A^{\prime i}({\bs x}, t)=\sum_\lambda \bs e_i^{\lambda} A_{\lambda}^\prime({\bs k, t})e^{i \bs k \bs x}, \quad a({\bs x}, t)= a({\bs k, t})e^{i \bs k \bs x},
\end{equation}
where $\bs e_i^\lambda$ is the $i$th component of the polarization vector of a photon with helicity $\lambda$, $A_\lambda(\bs k, t)$ and $A_\lambda^\prime(\bs k, t)$ are, respectively, the photon and dark photon amplitudes with helicity $\lambda$ while $a(\bs k, t)$ is the amplitude of the axion field. Consider ordinary and dark photons propagating along the observer's $z$ axis which points to the East, namely $\bs k=(0,0, k)$ and let $\bs B_e=B_e\hat{\bs n}, \bs B_e^\prime=B_e^\prime\hat{\bs n}^\prime$ where $\hat{\bs n}=[\cos(\Theta), \sin(\Theta)\cos(\Phi), \sin(\Theta)\sin(\Phi)]$ and $\hat{\bs n}^\prime=[\cos(\Theta^\prime), \sin(\Theta^\prime)\cos(\Phi^\prime), \sin(\Theta^\prime)\sin(\Phi^\prime)]$ are two generic direction unit vectors. Here $\Theta, \Theta^\prime$ are, respectively, the polar angles between magnetic fields $\bs B_e, \bs B_e^\prime$ and $x$ axis which points to North and $\Phi, \Phi^\prime$ are, respectively, the azimutal angles of $\bs B_e, \bs B_e^\prime$ with respect to $y$ axis which points outward. Now we can use the expansion \eqref{field-exp} in Eqs. \eqref{eq-A-ph} and then expand the operator $\partial_t^2+k^2=(-i\partial_t+k)(i\partial_t+k)$. After we look for solutions of the field amplitudes in the form $A_\lambda(k, t)=A_{k\lambda} (t)e^{-i\int\omega(t^\prime)\,dt^\prime}$,  $A_\lambda^\prime(k, t)=A_{k\lambda}^\prime (t)e^{-i\int\omega(t^\prime)\,dt^\prime}$, $a(k, t)=a_k (t)e^{-i\int\omega(t^\prime)\,dt^\prime}$ where $\omega$ is the particle energy and work in the WKB approximation, namely $\partial_t|A_{k\lambda}|\ll \omega |A_{k\lambda}|$, $\partial_t|A_{k\lambda}^\prime|\ll \omega |A_{k\lambda}^\prime|$, $\partial_t|a_{k}|\ll \omega |a_{k}|$. These approximations are valid when the time variation of the field amplitudes are much smaller than $\omega |A_{k\lambda}|, \omega |A_{k\lambda}^\prime|, \omega |a_{k}|$ or equivalently when variation in time of external magnetic fields are much smaller than photon/dark photon frequencies.

Now by acting on the fields with the term $(i\partial_t+k)$, which becomes $k+\omega$ while keeping untouched the second term  $(-i\partial_t+k)$, one can linearize Eq. \eqref{eq-A-ph} and get the following system of linear differential equations: 
\begin{equation}\label{schr-eq}
(i\partial_t-k)\Psi_k({t})\bs I+M \Psi_k(t)=0, 
\end{equation}
where $\bs I$ is the unit matrix, $\Psi_k(t)=(A_+, A_\times, A_+^\prime, A_\times^\prime, a)^\text{T}$ is a five component field and $M$ is the mixing matrix, which is given by
\begin{equation}\label{mixing-matrix}
 M=\begin{pmatrix}
  M_{+} & i M_F & 0 & 0 & i M_{a\gamma}^+ \\
-i M_F & M_{\times} & 0 & 0 & i M_{a\gamma}^\times\\
0 & 0 & M_+^\prime & i M_F^\prime & i M_{a\gamma}^{\prime +}\\
0 & 0 & -i M_F^\prime & M_\times^\prime & i M_{a\gamma}^{\prime\times}\\
-i M_{a\gamma}^+ & -i M_{a\gamma}^\times & -i M_{a\gamma}^{\prime +} & -i M_{a\gamma}^{\prime \times} & M_a
   \end{pmatrix}.\end{equation}
The photon states labeled with (+) are the linear polarization states which are parallel to the $y$ axis, namely $A_+\equiv A_y, A_+^\prime\equiv A_y^\prime$ while the states labelled with $(\times)$ are those which are parallel to the $x$ axis, $A_\times\equiv A_x, A_\times^\prime\equiv A_x^\prime$.  The elements of the mixing matrix $M$ are given by $M_a=-m_a^2/(\omega+k)$, $M_{a\gamma}^+=g_{a\gamma}\,\omega\,B_{e}\sin(\Theta)\cos(\Phi)/(\omega+k)$, $M_{a\gamma}^\times=g_{a\gamma}\,\omega\,B_{e}\cos(\Theta)/(\omega+k)$, $M_{a\gamma}^{\prime +}=g_{a\gamma}^\prime\,\omega\,B_{e}^\prime\sin(\Theta^\prime)\cos(\Phi^\prime)/(\omega+k)$, $M_{a\gamma}^{\prime\times}=g_{a\gamma}^\prime\,\omega\,B_{e}^\prime\cos(\Theta^\prime)/(\omega+k)$, $M_+=-\Pi^{22}/(\omega+k)$, $M_\times=-\Pi^{11}/(\omega+k), M_+^\prime=-\Pi^{\prime 22}/(\omega+k), M_\times^\prime=-\Pi^{\prime 11}/(\omega+k)$ and $M_F=i\,\Pi^{12}/(\omega+k), M_F^\prime=i\,\Pi^{\prime 12}/(\omega+k)$ are, respectively, the terms that include the Faraday effect\footnote{In the case when $\Phi=\pi/2$, the term $M_F$ includes solely the Faraday effect while for $\Phi\neq \pi/2$ it is a combination of Faraday and Cotton-Mouton effects, see Appendix \ref{appendix-B}. } in ordinary and dark media; see Appendix \ref{appendix-B} for the calculations of the matrix elements of $\Pi^{ij}$ in a magnetized plasma. The elements of $\Pi^{\prime ij}$ are formally the same as those of $\Pi^{ij}$ but with ordinary quantities that enter in $\Pi^{ij}$ replaced with those of the dark sector. Here $\omega$ is the total energy of the fields, namely $\omega=\omega_\gamma=\omega_{\gamma^\prime}=\omega_a$. In this work we assume that all particles participating in the mixing process are relativistic. In general, ordinary and dark photons are relativistic since the effects of the medium in generating an effective mass are very small. On the other hand, the axion can be either relativistic or not depending on its mass $m_a$. In the case when all particles participating in the mixing are relativistic, we can approximate $\omega+k\simeq 2k$ for $m_a\ll \omega$.


\section{Transition probability rates and Stokes parameters}
\label{sec:3}

The expressions for field amplitudes in \eqref{time-dep-sol} found by solving the equations of motion \eqref{schr-eq} are of extreme importance since we can derive very useful quantities such as the transition probabilities, phase shifts, the Stokes parameters etc. It is worth to stress that the expressions derived in  \eqref{time-dep-sol} are valid for arbitrary values of the angles $\Theta, \Theta^\prime, \Phi, \Phi^\prime$. In many situations is very convenient to have the expressions for the transition probabilities from one state into another in complete analogy with the case when the axion interacts with the ordinary sector only. However, the expressions for the transition probabilities for the case at hand are more complicated due to the interaction with the dark sector and due to the mixing of all ordinary and dark photon states with the axion state. This situation in principle can be simplified in the case when one knows the directions of ordinary and dark magnetic fields and then rotate the reference system in such a way as to ged rid of $M_F, M_F^\prime$ terms, and allow only one of the photon states to mix with the axion state. But typically the direction of the dark external magnetic field is not known, while for the ordinary external magnetic field there may be situations where its direction is known. In any case, in this section we derive general results without making any speculation about the magnetic fields directions.

The mixing of the axion with ordinary and dark photons makes possible the transition of ordinary photons into dark photons and vice-versa. The transition probabilities explicitly depend on the initial amplitude of fields at the initial time $t_\text{in}=0$. Assuming for example that initially $a(0)=0$, we get the following transition probability rates for $|A_\lambda(0)\rangle \rightarrow |A_+^\prime(t)\rangle$ (with $A_\lambda(0)=\delta_{\rho}^\lambda, A_+^\prime(0)=A_\times^\prime(0)=0$) and $|A_\lambda^\prime(0)\rangle \rightarrow |A_+(t)\rangle$ (with $A_+(0)=A_\times(0)=0, A_\lambda^\prime (0)=\delta_\rho^\lambda$ where $\rho=+, \times$)
\begin{equation}
\begin{gathered}
P[|A_+(0)\rangle \rightarrow |A_+^\prime(t)\rangle]  =  \left| \int_0^{t}\int_0^{t^\prime}\,dt^{\prime}dt^{\prime\prime}\,M_{a\gamma}^{\prime +}(t^{\prime})M_{a\gamma}^{+}(t^{\prime\prime}) e^{-i\left(\Delta M_1^\prime(t^\prime)-\Delta M_1(t^{\prime\prime})\right)}\right|^2,  \nonumber\\ 
P[|A_\times (0)\rangle\rightarrow |A_+^\prime(t)\rangle]  = \left| \int_0^{t}\int_0^{t^\prime}\,dt^{\prime}dt^{\prime\prime}\,M_{a\gamma}^{\prime +}(t^{\prime})M_{a\gamma}^\times(t^{\prime\prime}) e^{-i\left(\Delta M_1^\prime(t^\prime)-\Delta M_2(t^{\prime\prime})\right)}\right|^2, \nonumber
\end{gathered}
\end{equation}

\begin{equation}\label{P-A-Ap}
\begin{gathered}
P[A_+^\prime(0)\rightarrow A_+(t)]  =  \left| \int_0^{t}\int_0^{t^\prime}\,dt^{\prime}dt^{\prime\prime}\,M_{a\gamma}^+(t^{\prime})M_{a\gamma}^{\prime +}(t^{\prime\prime}) e^{-i\left(\Delta M_1^\prime(t^\prime)-\Delta M_1(t^{\prime\prime})\right)} \right|^2, \\
P[A_\times^\prime(0)\rightarrow A_+(t)]  =  \left| \int_0^{t}\int_0^{t^\prime}\,dt^{\prime}dt^{\prime\prime}\,M_{a\gamma}^+(t^{\prime})M_{a\gamma}^{\prime \times}(t^{\prime\prime}) e^{-i\left(\Delta M_1^\prime(t^\prime)-\Delta M_2(t^{\prime\prime})\right)} \right|^2.
\end{gathered}
\end{equation}

On the other hand the transition probabilities from $|A_\lambda(0)\rangle \rightarrow |a(t)\rangle$ (with $A_\lambda(0)=\delta_{\rho}^\lambda, A_+^\prime(0)=A_\times^\prime(0)=0$) and $|A_\lambda^\prime(0)\rangle \rightarrow |a(t)\rangle$ (with $A_+(0)=A_\times(0)=0, A_\lambda^\prime (0)=\delta_\rho^\lambda$) are, respectively, given by
\begin{equation}
\begin{gathered}
P[|A_+(0)\rangle \rightarrow |a(t)\rangle]  =  \left|  \int_0^t dt^\prime\,M_{a\gamma}^+(t^\prime) e^{i\Delta M_1(t^\prime)} - \int_0^{t}\int_0^{t^\prime}\,dt^{\prime}dt^{\prime\prime}\,M_{a\gamma}^{\times}(t^{\prime})M_F(t^{\prime\prime}) e^{i\left(\Delta M_2(t^\prime)+\Delta M(t^{\prime\prime})\right)} \right|^2, \nonumber\\
P[|A_\times(0)\rangle \rightarrow |a(t)\rangle]  =  \left|  \int_0^t dt^\prime\,M_{a\gamma}^\times(t^\prime) e^{i\Delta M_2(t^\prime)} + \int_0^{t}\int_0^{t^\prime}\,dt^{\prime}dt^{\prime\prime}\,M_{a\gamma}^{+}(t^{\prime})M_F(t^{\prime\prime}) e^{i\left(\Delta M_1(t^\prime)-\Delta M(t^{\prime\prime})\right)} \right|^2, \nonumber
\end{gathered}
\end{equation}
\begin{equation}\label{P-A-a}
\begin{gathered}
P[|A_+^\prime(0)\rangle \rightarrow |a(t)\rangle]  =  \left|  \int_0^t dt^\prime\,M_{a\gamma}^{\prime +}(t^\prime) e^{i\Delta M_1^\prime(t^\prime)} - \int_0^{t}\int_0^{t^\prime}\,dt^{\prime}dt^{\prime\prime}\,M_{a\gamma}^{\prime \times}(t^{\prime})M_F^\prime(t^{\prime\prime}) e^{i\left(\Delta M_2^\prime(t^\prime)+\Delta M^\prime(t^{\prime\prime})\right)}\right|^2, \\
P[|A_\times^\prime(0)\rangle \rightarrow |a(t)\rangle]  =  \left| \int_0^t dt^\prime\,M_{a\gamma}^{\prime \times}(t^\prime) e^{i\Delta M_2^\prime(t^\prime)} + \int_0^{t}\int_0^{t^\prime}\,dt^{\prime}dt^{\prime\prime}\,M_{a\gamma}^{\prime +}(t^{\prime})M_F^\prime(t^{\prime\prime}) e^{i\left(\Delta M_1^\prime(t^\prime)-\Delta M^\prime(t^{\prime\prime})\right)}  \right|^2. 
\end{gathered}
\end{equation}
We may note that in \eqref{P-A-a} there is no contribution of the dark sector to the transition probabilities $P[|A_\lambda(0)\rangle\rightarrow |a(t)\rangle]$ to second order in perturbation theory and there is no contribution, to second order, of the ordinary sector to the transition probabilities $P[|A_\lambda^\prime(0)\rangle \rightarrow |a(t)\rangle]$. The contributions of, respectively, the dark and ordinary sectors in \eqref{P-A-a} start from the third order of iteration. 

The transition probability rates calculated in \eqref{P-A-a} are very important in those situations where one is not interested directly in the polarization state of the light. However, there might be situations where one is mostly interested in the polarization state of light and consequently the transition probability rates are not useful in this case. Instead, the Stokes parameters are those quantities which give us information as regards the polarization state of the  light. They are usually defined in terms of the electric field components $\bs E_i$ in a cartesian reference system but here we define them in terms of vector potential components $\bs A_i$ as follows:
\begin{eqnarray}\label{Stokes-par}
I_\gamma(t) & \equiv & |A_\times(t)|^2+|A_+(t)|^2, \quad Q(t) \equiv |A_\times(t)|^2-|A_+(t)|^2,\nonumber\\
 U(t) & \equiv & 2\,\text{Re}\left\{A_\times(t)A_+^*(t) \right\}, \quad  V(t) \equiv -2\,\text{Im}\left\{A_\times(t)A_+^*(t) \right\}.
 \end{eqnarray}

Now by writing the field amplitudes as show in \eqref{amplitudes}, using the expressions for $|A_+(t)|^2$ and $|A_\times(t)|^2$ derived in Appendix \ref{Appendix-A}, using the definitions of the Stokes parameters in \eqref{Stokes-par} and then after lengthy calculations, we get the following expressions in the case when $a(0)=0$
\begin{equation}\label{ph-int}
\begin{gathered}
I_\gamma(t)= \left(|I_2(t)|^2 + |I_6(t)|^2\right) |A_\times(0)|^2 + \left(|I_1(t)|^2 + |I_5(t)|^2\right) |A_+(0)|^2  +  \left(|I_3(t)|^2 + |I_7(t)|^2\right) |A_+^\prime(0)|^2 + \left(|I_4(t)|^2 \right. \\ \left. +  |I_8(t)|^2\right) |A_\times^\prime(0)|^2 - 2\,\text{Re}\left\{\left[I_1(t)I_2^*(t)+I_5(t)I_6^*(t)\right]A_+(0)A_\times^*(0) + \left[I_1(t)I_3^*(t)-I_5(t)I_7^*(t)\right]A_+(0)A_+^{\prime *}(0) \right. \\ \left. + \left[I_1(t)I_4^*(t)- I_5(t)I_8^*(t) \right]A_+(0)A_\times^{\prime *}(0)\right\} + 2\,\text{Re}\left\{\left[I_2(t)I_3^*(t) - I_6(t)I_7^*(t)\right] A_\times(0)A_+^{\prime *}(0) + \right. \\ \left.  \left[I_2(t)I_4^*(t) - I_6(t)I_8^*(t) \right] A_\times(0)A_\times^{\prime *}(0) + \left[I_3(t)I_4^*(t) + I_7(t)I_8^*(t)\right]A_+^\prime(0)A_\times^{\prime *}(0)\right\}, \\
\\
Q(t)= \left(|I_6(t)|^2 - |I_2(t)|^2 \right) |A_\times(0)|^2 + \left(|I_5(t)|^2 - |I_1(t)|^2 \right) |A_+(0)|^2 +\left(|I_7(t)|^2 - |I_3(t)|^2\right) |A_+^\prime(0)|^2 +  
\left(|I_8(t)|^2 \right. \\ \left. -  |I_4(t)|^2\right) |A_\times^\prime(0)|^2 - 2\,\text{Re}\left\{\left[I_5(t)I_6^*(t) - I_1(t)I_2^*(t)\right]A_+(0)A_\times^*(0) - \left[I_1(t)I_3^*(t)+I_5(t)I_7^*(t)\right]A_+(0)A_+^{\prime *}(0) \right. \\ \left. - \left[I_1(t)I_4^*(t) + I_5(t)I_8^*(t) \right]A_+(0)A_\times^{\prime *}(0)\right\} - 2\,\text{Re}\left\{\left[I_6(t)I_7^*(t) + I_2(t)I_3^*(t)\right] A_\times(0)A_+^{\prime *}(0) - \right. \\ \left.  \left[I_2(t)I_4^*(t) + I_6(t)I_8^*(t) \right] A_\times(0)A_\times^{\prime *}(0) - \left[I_7(t)I_8^*(t) - I_3(t)I_4^*(t)\right]A_+^\prime(0)A_\times^{\prime *}(0)\right\},\\
\\
U(t) = 2\,\text{Re}\left\{A_\times(t)A_+^*(t) \right\}, \quad  V(t) = -2\,\text{Im}\left\{A_\times(t)A_+^*(t) \right\},
\end{gathered}
\end{equation}
where
\begin{equation}
\begin{gathered}
A_\times(t)A_+^*(t)=-I_6(t)I_2^*(t) |A_\times(0)|^2-I_5(t)I_1^*(t) |A_+(0)|^2 + I_7(t)I_3^*(t) |A_+^\prime(0)|^2 + I_8(t)I_4^*(t) |A_\times^\prime(0)|^2 \nonumber\\
+ I_5(t)I_2^*(t) A_+(0)A_\times^*(0)+ I_5(t)I_3^*(t) A_+(0)A_+^{\prime *}(0) + I_5(t)I_4^*(t) A_+(0)A_\times^{\prime *}(0) + I_6(t)I_1^*(t) A_\times(0)A_+^{*}(0)\nonumber \\
- I_6(t)I_3^*(t) A_\times(0)A_+^{\prime *}(0) - I_6(t)I_4^*(t) A_\times(0)A_\times^{\prime *}(0) - I_7(t)I_1^*(t) A_+^\prime(0)A_+^{*}(0) + I_7(t)I_2^*(t) A_+^\prime(0)A_\times^{*}(0)\nonumber \\
+ I_7(t)I_4^*(t) A_+^\prime(0)A_\times^{\prime *}(0) - I_8(t)I_1^*(t) A_\times^\prime(0)A_+^{*}(0) +
I_8(t)I_2^*(t) A_\times^\prime(0)A_\times^{*}(0) + I_8(t)I_3^*(t) A_\times^\prime(0)A_+^{\prime *}(0). 
\end{gathered}
\end{equation}
It is worth to stress that the expressions for the Stokes parameters in \eqref{ph-int} are valid for any direction of photon propagation with respect to the external ordinary and dark magnetic fields, namely for any values of the angles $\Theta, \Theta^\prime, \Phi, \Phi^\prime$. In addition, is quite straightforward to see from the definitions of $I_1(t)$ and $I_6(t)$ that $I_\gamma(t)= I_\gamma(0) + \text{other terms}$, where $I_\gamma(0)= |A_\times(0)|^2 + |A_+(0)|^2$ and the other terms can have either signs.

\section{Effects on ordinary and dark CMBs}
\label{sec:4}

The model of photon-axion-dark photon mixing which we discussed above may have several applications. However, before applying it to a concrete example, is important to recall that in order to have photon-dark photon mixing there must necessarily exist an external dark magnetic field in addition to the ordinary one. Obviously, laboratory experiments looking for axions and dark photons are ruled out since one can generate in the laboratory an ordinary magnetic field but not a dark magnetic field. This fundamental observation tells us that we must look for this effect elsewhere, possibly in astrophysical or cosmological situations where ordinary and dark magnetic fields coexist. 

One possibility to apply our model is in cosmology or, more precisely, in the context of CMB physics. Indeed, as already mentioned in Sect. \ref{sec:1}, based on the concept of M-symmetry one would expects that both sectors have similar cosmological evolution and same microphysics. In order to avoid any conflict with the BBN, the two sectors must have different initial conditions and different temperatures $T\neq T^\prime$ at the reheating epoch \cite{Berezhiani:1995am}. The BBN bound on the number of effective neutrino species puts very stringent limit on the temperature of the dark CMB which must be $T^\prime<0.64\, T$ \cite{Berezhiani:2000gw} where $T$ is the temperature of the ordinary CMB.

Since ordinary and dark CMBs evolve with different temperatures and because they do not come in thermal equilibrium with each other, one would also expect the dark CMB to experience a decoupling epoch which happens to be slightly earlier than the ordinary decoupling epoch. Therefore one would also expect that there must exist a large scale dark magnetic field complementary to the ordinary large-scale magnetic field. Consequently, we would have two CMBs, one ordinary and one dark, where each of them interacts with its respective large-scale magnetic field. Based on this assertion, we may use our earlier formalism of photon-axion-dark photon mixing in order to study the effects which the dark CMB has on the ordinary CMB.

Here we illustrate one possible effect that the dark CMB has on the ordinary CMB due to photo-axion-dark photon mixing, namely it generates for example a temperature anisotropy. In order to calculate this effect, first we must recall that according to the standard cosmology, the temperature anisotropy of the ordinary CMB is essentially generated at the decoupling time or afterwards due to several processes. Here we consider the case where the CMB acquires a temperature anisotropy starting from the decoupling epoch and it continues evolving until at the present epoch. Therefore, we must study the evolution of the temperature anisotropy in this time interval and consequently all quantities of interest would evolve in time. So, based on this observation we must use our time dependent formalism developed in Appendix \ref{Appendix-A}.

Consider now two observation directions in the ordinary sector, namely one parallel to the ordinary external magnetic field and one perpendicular to it. Here we are supposing that we know the direction of $\bs B_e$ but the direction of $\bs B_e^\prime$ is supposed to be not known. The generalization when even $\bs B_e$ is not known and might change in time randomly can easily be taken into account by averaging the ordinary sector quantities over $\Phi$ and $\Theta$. The accuracy between considering a fixed direction of observation in the ordinary sector instead of averaging over $\Phi$ and $\Theta$ is expected to be within an order of magnitude. In addition, in this section we assume that axions are initially absent at the ordinary and dark decoupling epochs, namely $a(0)=0$.

In the direction perpendicular to the ordinary external magnetic field the intensity of the CMB would be\footnote{The proportionality factor in $I_\gamma^\perp$ and $I_\gamma^{||}$ in general is a factor which takes into account the dilution of the particle number density in an expanding universe and it cancels out in the final result. } $I_\gamma^\perp(t_0)\propto (|A_+(t_0)|^2+ |A_\times(t_0)|^2)$ where $A_+(t)$ and $A_\times(t)$ are given in expression \eqref{time-dep-sol}. In this direction, namely $\Theta=0$ only the quantities $M_F(t_0)\propto \sin(\Phi)\sin(\Theta)$ and $M_{a\gamma}^+(t_0)\propto \cos(\Phi)\sin(\Theta)$ in the mixing matrix $M$ are zero. This implies also that in expressions \eqref{integrals},  $I_2(t_0)=I_3(t_0)=I_4(t_0)=I_5(t_0)=0$ while the other integrals are different from zero. 

In the direction parallel to the ordinary external magnetic field $\Theta=\pi/2$, consider also that $\bs B_e$ is in the $xz$ plane with $\Phi=\pi/2$. For this configuration, we have only $M_{a\gamma}^+(t_0)\propto \cos(\Phi)\sin(\Theta)=0$ and $M_{a\gamma}^\times(t_0)\propto \cos(\Theta)=0$. In this case there is not generation of axions since ordinary photons do not mix with the axion because the ordinary and dark sectors are decoupled from each other. In this case, the intensity of the states $A_+$ and $A_\times$ changes only due to the Faraday effect. However, it is well known that the Faraday effect does not change the total intensity of light but only its polarization state. This fact can easily be verified by considering the case where only the Faraday effect is present in the ordinary photon mixing matrix $M$ with $M_+=M_\times$. Consequently, the intensity of the light parallel to the external magnetic field would be $I_\gamma^{||}(t_0)\propto  (|A_+(0)|^2+ |A_\times(0)|^2)$, namely it is equal to the intensity of light in an unperturbed universe.

Consider the CMB in a thermal state at the ordinary decoupling time where its intensity is given by the black body formula. In this case one can derive the following relation, to first order, between the differential intensity and temperature changes: $\delta I_\gamma(t_0)/I_\gamma(t_0)\simeq [xe^x/(e^x-1)]\delta T_0/T_0$ where $T_0$ is the present day temperature of the CMB averaged over all directions in the sky and $x=2\pi\nu_0/T_0$ with $\nu_0$ being the CMB frequency at the present epoch. Since changes in the ordinary CMB intensity or temperature are very small for two given observation directions, we would have 
\begin{equation}\label{int-diff}
\frac{I_\gamma^\perp(t_0)-I_\gamma^{||}(t_0)}{I_\gamma(t_0)}\simeq \frac{\delta I_\gamma(t_0)}{I_\gamma(t_0)}.
\end{equation}

In the direction perpendicular to the ordinary magnetic field, the intensity of ordinary photons is given by
\begin{align}\label{par-In}
I_\gamma^\perp(t_0) & \propto  \left[I_\gamma(0)+\left(|\mathcal I(t_0)|^2-2\text{Re}\{\mathcal I(t_0)\}\right)|A_\times(0)|^2+|I_7(t_0)|^2|A_+^\prime(0)|^2+|I_8(t_0)|^2|A_\times^\prime(0)|^2 - \right.\nonumber \\ 
& \left. 2\,\text{Re}\left\{I_6(t_0)I_7^*(t_0) A_\times(0)A_+^{\prime *}(0) + I_6(t_0)I_8^*(t_0) A_\times(0)A_\times^{\prime *}(0) - I_7(t_0)I_8^*(t_0) A_+^\prime(0)A_\times^{\prime *}(0) \right\}\right],
\end{align}
where we used the expression \eqref{ph-int} for the perpendicular propagation with respect to the ordinary magnetic field with $I_2(t_0)=I_3(t_0)=I_4(t_0)=I_5(t_0)=0$ and have defined 
\begin{align}
\mathcal I(t_0) &\equiv \int_0^{t_0}\int_0^{t^\prime}\,dt^{\prime}dt^{\prime\prime} M_{a\gamma}^\times(t^\prime)M_{a\gamma}^\times(t^{\prime\prime})e^{-i\left(\Delta M_2(t^\prime)-\Delta M_2(t^{\prime\prime})\right)}. \nonumber
\end{align}
Now by using the fact that for parallel propagation the intensity of ordinary photons is $I_\gamma^{||}(t_0) \propto (|A_+(0)|^2 + |A_\times(0)|^2)$ and using the expression \eqref{par-In}, we get
\begin{align}\label{delta-In}
I_\gamma^\perp(t_0)-I_\gamma^{||}(t_0) & \propto \left[ \left(|\mathcal I(t_0)|^2-2\text{Re}\{\mathcal I(t_0)\}\right)|A_\times(0)|^2+|I_7(t_0)|^2|A_+^\prime(0)|^2+|I_8(t_0)|^2|A_\times^\prime(0)|^2 - \right. \nonumber \\ 
& \left. 2\,\text{Re}\left\{I_6(t_0)I_7^*(t_0) A_\times(0)A_+^{\prime *}(0) + I_6(t_0)I_8^*(t_0) A_\times(0)A_\times^{\prime *}(0) - I_7(t_0)I_8^*(t_0) A_+^\prime(0)A_\times^{\prime *}(0) \right\} \right].
\end{align}

At this point we make the assumption that at the ordinary decoupling time, the dark CMB is roughly speaking in a thermal state. In addition, by averaging over the polarization states at the initial time $t_\text{in}=0$, which we choose to coincide with the ordinary decoupling time, we get $\langle |A_+(0)|^2\rangle=\langle |A_\times(0)|^2\rangle =(1/2)I_\gamma(0)$, $\langle |A_+^\prime(0)|^2\rangle=\langle |A_\times^\prime(0)|^2\rangle =(1/2)I_{\gamma}^\prime(0)$ where the symbol $\langle(.)\rangle$ expresses the average value over the initial polarization states of ordinary and dark photons. Moreover, we assume that the mixed terms $\langle A_\times(0)A_\times^{\prime *}(0)\rangle=\langle A_\times(0)A_+^{\prime *}(0)\rangle=\langle A_+^\prime(0)A_\times^{\prime *}(0)\rangle = 0$.
Making use of these considerations in expression \eqref{delta-In}, we get the following relation for the averaged value over the initial polarization states of the fractional change of the ordinary photon intensity: 
\begin{equation}\label{int-diff-1}
\frac{\langle I_\gamma^\perp (t_0)-I_\gamma^{||}(t_0)\rangle}{\langle I_\gamma(t_0)\rangle }=\frac{\left(|\mathcal I(t_0)|^2-2\text{Re}\{\mathcal I(t_0)\}\right)\langle|A_\times(0)|^2\rangle+| I_7(t_0)|^2 \langle|A_+^\prime(0)|^2\rangle + | I_8(t_0)|^2 \langle|A_\times^\prime(0)|^2 \rangle}{\langle |A_+(0)|^2\rangle+\langle|A_\times(0)|^2\rangle},
\end{equation}
where $I_\gamma(0)$ is the photon intensity in an unperturbed universe which in our case is equal to $I_\gamma^{||}(0)$.
By expressing all average values of amplitudes square in \eqref{int-diff-1} in terms of the photon intensities and using the expression \eqref{int-diff}, we get the following expression which relates the CMB temperature anisotropy between two directions at $90^\circ$ in the sky with the photon and dark photon intensities 
\begin{equation}\label{temp-int}
\frac{\left(|\mathcal I(t_0)|^2-2\text{Re}\{\mathcal I(t_0)\}\right) I_\gamma(0)+\left(|I_7(t_0)|^2 + |I_8(t_0)|^2 \right) I_{\gamma}^\prime(0)}{2\,I_\gamma(0)}=\left(\frac{xe^x}{e^x-1}\right) \left.\frac{\delta T_0}{T_0}\right|_{90^\circ}.
\end{equation}

We must stress that the expression \eqref{temp-int} has been derived by using the expressions of fields up to the second order in the perturbation theory where $M_1(t)$ has been considered as a perturbation matrix with respect to $M_0(t)$. In addition, we considered the ordinary photon intensity difference between the direction parallel and perpendicular with respect to the ordinary external magnetic field. The first term on the left hand side of \eqref{temp-int} reflects the change in the photon intensity due to photon-axion mixing which results in a decrease of the ordinary photon intensity while the second term is the contribution of the conversion of the dark photons into ordinary  photons, namely it is a gain term. In general, the relative magnitude of the two terms would depend on several parameters and one would  expect the photon-axion contribution to be the dominant term. It is worth also to stress that the expression \eqref{temp-int} is valid for arbitrary direction of the dark magnetic field $\bs B_{e}^\prime$ with respect to the direction of observation. The dependence of $|I_7(t_0)|^2$ and $|I_8(t_0)|^2$ in \eqref{temp-int} on the angles $\Theta^\prime$ and $\Phi^\prime$ is straightforwardly averaged out in the case when the angles $\Theta^\prime$ and $\Phi^\prime$ are independent on the time. In the case when $\Theta^\prime$ and $\Phi^\prime$ depend on the time, one can still average out the contribution of the dark sector by assuming $\Theta^\prime$ and $\Phi^\prime$ as random functions of the time.

\section{Conclusions}
\label{sec:5}

In this work we proposed and studied the effect of the photo-axion-dark photon mixing in external ordinary and dark magnetic fields. As a consequence of this mixing, dark photons can interact with the ordinary photons via the same axion field. Then we solved equations of motion for time depended mixing matrix where perturbative solutions for the photon, dark photon and axion fields have been found. The derived results can be applied in the cases when ordinary and dark photons propagate through time dependent magnetized media such as those present in cosmological situations. With the introduction of the dark photon in the mixing problem, the usual expressions for the photon-axion transition probability rates, Stokes parameters etc., get modified. This fact could have a significant impact in those situations where an external dark magnetic field is present  and one needs to know the magnitude of these quantities in order to compare them with experimentally measurable quantities.

Our results have been derived by neglecting the weak gravitational interaction between the two sectors and considered their interaction only through the same axion field. In our model ordinary and dark photons interact solely through the axion field. In principle, one could also include in the interaction Lagrangian density a kinetic mixing between photons and dark photons, namely $\mathcal L_I\propto \epsilon F_{\mu\nu}F^{\prime\mu\nu}$, which is not forbidden by the M-symmetry. The inclusion of such term is only optional and can easily be accommodated in our formalism.

In order for the photon-axion-dark photon mixing to work there must coexist in the same place and time both ordinary and dark magnetic fields. The only possibility to apply this mixing, happens to be in astrophysical and cosmological situations. In this work we applied our mechanisms in the context of CMB physics and showed as a matter of example that the photon-axion-dark photon mixing would generates a CMB temperature anisotropy at the ordinary post decoupling epoch. The same effect would also generates polarization of the CMB as is evident from the expressions of the Stokes parameters in \eqref{ph-int}. In an astrophysical situation, our model could be used in order to calculate the generated flux of photons in ordinary and dark magnetic fields by dark stars and other dark objects which emit dark photons, where the generated flux might contributes to well know galactic and/or extragalactic backgrounds.

With respect to the case of photon-axion mixing, our model has additional free parameters. Indeed, by a close inspection of the expression \eqref{coup-cons} we may observe that the coupling constants are related with each other through $f_a$, namely the coupling constants are proportional to each other. The proportionality term is a combination of $z$ and $z^\prime$ where the former is usually known while the latter is less known. If both $z, z^\prime$ are known, the number of independent parameters is either $m_a$ or $g_{a\gamma}$ or $g_{a\gamma}^\prime$ similarly as in the case of the photon-axion mixing. Additional implicit parameters of our model essentially do appear in the index of refraction of dark photons which usually contains the plasma frequency which is related to the number density of the free dark electrons and to the amplitude of the dark magnetic field.

In the context of the CMB physics, our model can be applied to constrain the parameter space of axions which are essentially either the coupling constant to photons and dark photons or its mass. Indeed, the expression \eqref{temp-int} can be used to limit/constrain the axion parameter space and/or the magnetic field amplitudes based on the known value of the amplitude of the CMB temperature anisotropy. On the other hand, if one knows the values of the parameters which enter in \eqref{temp-int}, one can estimates which is the contribution of photon-axion-dark photon mixing to the CMB temperature anisotropy. The presence of the ordinary large-scale magnetic field generates a CMB temperature anisotropy by itself, so, the result \eqref{temp-int} gives only the contribution of the photon-axion-dark photon mixing to the total CMB temperature anisotropy. On the other hand, even though we studied for simplicity only the effects of the photon-axion-dark photon mixing on the CMB temperature anisotropy, additional limits/constraints can be inferred from the present limits on the CMB polarization. Indeed, our model generates also birefringence and dichroism effects on the CMB, namely it generates an elliptic polarization with non zero Stokes parameters $Q(t), U(t)$ and $V(t)$, as one can observe from the expression \eqref{ph-int}.

\vspace{1cm}

\hrulefill

 {\bf{AKNOWLEDGMENTS}}:
I would like to thank Zurab Berezhiani for very interesting discussions and for letting me know about his work on mirror world. This work is supported by the Russian Science Foundation Grant no. 16-12-10037. I also would like to  thank LNGS for the support received through the fellowship POR 2007-2013 `Sapere e Crescita' where part of this research was conducted.

\hrulefill

\vspace{1cm}

  \appendix
  \section{Perturbative solutions of equations of motion and photon field amplitudes}
\label{Appendix-A}
\numberwithin{equation}{section}

In the case when the field mixing matrix $M$ is time dependent, in general is not possible to find exact closed solutions but one might attempt to look for solutions by using the perturbation theory. In this regard, we can split the mixing matrix in the following way\footnote{Here we are simply treating the case when $M_1(t)$ is considered a small perturbation since in most practical cases the magnitude of elements of $M_0(t)$ are bigger than those of $M_1(t)$ independently on the values of the angles $\Theta, \Phi, \Theta^\prime, \Phi^\prime$.} $M(t)=M_0(t)+M_1(t)$ where $M_0(t)=\text{diag}\left[M_+(t), M_\times(t), M_+^\prime(t), M_\times^\prime(t), M_a(t)\right]$ is a diagonal matrix and $M_1(t)$ is a small perturbation matrix given by
\begin{equation}\nonumber
M_1(t)=\begin{pmatrix}
 0 & i M_F & 0 & 0 & i M_{a\gamma}^+ \\
-i M_F & 0 & 0 & 0 & i M_{a\gamma}^\times\\
0 & 0 & 0& i M_F^\prime & i M_{a\gamma}^{\prime +}\\
0 & 0 & -i M_F^\prime & 0 & i M_{a\gamma}^{\prime\times}\\
-i M_{a\gamma}^+ & -i M_{a\gamma}^\times & -i M_{a\gamma}^{\prime +} & -i M_{a\gamma}^{\prime\times} & 0
   \end{pmatrix}.
      \end{equation}
Now is convenient to move to the interaction picture by defining $\Psi_\text{int}(t)=U^\dagger(t) \Psi(t)$ (where we dropped the index $k$ on $\Psi$ for simplicity) and $M_\text{int}(t)=U^\dagger(t) M_1(t) U(t)$ where $U(t)=\exp[-i\int_0^t dt^\prime \left(k(t^\prime)\bs I- M_0(t^\prime)\right)]$. In the interaction picture, Eq. \eqref{schr-eq} becomes $i\partial_t\Psi_\text{int}(t)=M_\text{int}(t)\Psi_\text{int}(t)$. By using the standard iterative procedure, we find the following perturbative solution for $\Psi_\text{int}(t)$ to the first and second order in the perturbation matrix $M_\text{int}(t)$
\begin{equation}
\Psi_\text{int}^{(1)}(t)=-i\int_0^t dt^\prime\, M_\text{int}(t^\prime)\Psi(0), \quad \Psi_\text{int}^{(2)}(t)=-\int_0^t \int_0^{t^\prime} dt^\prime\, dt^{\prime\prime}\,  M_\text{int}(t^\prime)\,M_\text{int}(t^{\prime\prime})\Psi(0),
\end{equation}
where $\Psi_\text{int}^{(0)}(t)=\Psi(0)$, $\Psi_\text{int}(t)=\Psi_\text{int}^{(0)}(t)+\Psi_\text{int}^{(1)}(t)+\Psi_\text{int}^{(2)}(t)+ \text{higher order terms}$, and we have chosen for simplicity the initial time $t_\text{in}=0$. Performing several algebraic operations, we get the following solutions for the field amplitudes up to the second order in perturbation theory in the Schr\"{o}dinger  picture
\begin{equation}\nonumber
\begin{gathered}
A_\text{+}(t) =\left[1- \int_0^{t}\int_0^{t^\prime}\,dt^{\prime}dt^{\prime\prime}\,M_{a\gamma}^+(t^{\prime})M_{a\gamma}^+(t^{\prime\prime}) e^{-i\left(\Delta M_1(t^\prime)-\Delta M_1(t^{\prime\prime})\right)}   - \int_0^{t}\int_0^{t^\prime}\,dt^{\prime}dt^{\prime\prime}\,M_F(t^{\prime})M_F(t^{\prime\prime})\times\right.\\ \left.  e^{-i\left(\Delta M(t^\prime)-\Delta M(t^{\prime\prime})\right)} \right]e^{-i\tilde M_+(t)}A_+(0) + \left[  \int_0^t dt^\prime\,M_F(t^\prime) e^{-i\Delta M(t^\prime)} - \int_0^{t}\int_0^{t^\prime}\,dt^{\prime}dt^{\prime\prime}\,M_{a\gamma}^+(t^{\prime})M_{a\gamma}^\times(t^{\prime\prime})\times\right. \\ \left. e^{-i\left(\Delta M_1(t^\prime)-\Delta M_2(t^{\prime\prime})\right)} \right]\,e^{-i\tilde M_+(t)}\,A_\times(0) - \left[ \int_0^{t}\int_0^{t^\prime}\,dt^{\prime}dt^{\prime\prime}\,M_{a\gamma}^+(t^{\prime})M_{a\gamma}^{\prime +}(t^{\prime\prime}) e^{-i\left(\Delta M_1^\prime(t^\prime)-\Delta M_1(t^{\prime\prime})\right)} \right]e^{-i\tilde M_+(t)}A_+^\prime(0)\\   
- \left[ \int_0^{t}\int_0^{t^\prime}\,dt^{\prime}dt^{\prime\prime}\,M_{a\gamma}^+(t^{\prime})M_{a\gamma}^{\prime\times}(t^{\prime\prime}) e^{-i\left(\Delta M_1(t^\prime)-\Delta M_2^\prime(t^{\prime\prime})\right)} \right]e^{-i\tilde M_+(t)}A_\times^\prime(0) + \left[  \int_0^t dt^\prime\,M_{a\gamma}^+(t^\prime) e^{-i\Delta M_1(t^\prime)} \right. \\ \left. + \int_0^{t}\int_0^{t^\prime}\,dt^{\prime}dt^{\prime\prime}\,M_F(t^{\prime})M_{a\gamma}^\times(t^{\prime\prime}) e^{-i\left(\Delta M(t^\prime)+\Delta M_2(t^{\prime\prime})\right)} \right]e^{-i\tilde M_+(t)}a(0),
\end{gathered}
\end{equation}
\begin{equation}\nonumber
\begin{gathered}
A_\times(t) =-\left[  \int_0^t dt^\prime\,M_F(t^\prime) e^{i\Delta M(t^\prime)} + \int_0^{t}\int_0^{t^\prime}\,dt^{\prime}dt^{\prime\prime}\,M_{a\gamma}^\times(t^{\prime})M_{a\gamma}^+(t^{\prime\prime}) e^{-i\left(\Delta M_2(t^\prime)-\Delta M_1(t^{\prime\prime})\right)} \right]\,e^{-i\tilde M_\times(t)}\,A_+(0)\\
 +\left[1- \int_0^{t}\int_0^{t^\prime}\,dt^{\prime}dt^{\prime\prime}\,M_{a\gamma}^\times(t^{\prime})M_{a\gamma}^\times(t^{\prime\prime}) e^{-i\left(\Delta M_2(t^\prime)-\Delta M_2(t^{\prime\prime})\right)}   - \int_0^{t}\int_0^{t^\prime}\,dt^{\prime}dt^{\prime\prime}\,M_F(t^{\prime})M_F(t^{\prime\prime})\times\right.\\ \left.  e^{i\left(\Delta M(t^\prime)-\Delta M(t^{\prime\prime})\right)} \right]e^{-i\tilde M_\times(t)}A_\times(0)- \left[ \int_0^{t}\int_0^{t^\prime}\,dt^{\prime}dt^{\prime\prime}\,M_{a\gamma}^\times(t^{\prime})M_{a\gamma}^{\prime +}(t^{\prime\prime}) e^{-i\left(\Delta M_2(t^\prime)-\Delta M_1^\prime(t^{\prime\prime})\right)} \right]e^{-i\tilde M_\times(t)}A_+^\prime(0)\\   
- \left[ \int_0^{t}\int_0^{t^\prime}\,dt^{\prime}dt^{\prime\prime}\,M_{a\gamma}^\times(t^{\prime})M_{a\gamma}^{\prime\times}(t^{\prime\prime}) e^{-i\left(\Delta M_2(t^\prime)-\Delta M_2^\prime(t^{\prime\prime})\right)} \right]e^{-i\tilde M_\times(t)}A_\times^\prime(0) + \left[  \int_0^t dt^\prime\,M_{a\gamma}^\times(t^\prime) e^{-i\Delta M_2(t^\prime)} \right. \\ \left. - \int_0^{t}\int_0^{t^\prime}\,dt^{\prime}dt^{\prime\prime}\,M_F(t^{\prime})M_{a\gamma}^+(t^{\prime\prime}) e^{i\left(\Delta M(t^\prime)-\Delta M_1(t^{\prime\prime})\right)} \right]e^{-i\tilde M_\times(t)}a(0),
\end{gathered}
\end{equation}
\begin{equation}\nonumber
\begin{gathered}
A_+^\prime(t) =- \left[ \int_0^{t}\int_0^{t^\prime}\,dt^{\prime}dt^{\prime\prime}\,M_{a\gamma}^{\prime +}(t^{\prime})M_{a\gamma}^{+}(t^{\prime\prime}) e^{-i\left(\Delta M_1^\prime(t^\prime)-\Delta M_1(t^{\prime\prime})\right)} \right]e^{-i\tilde M_+^\prime(t)}A_+(0) - \left[ \int_0^{t}\int_0^{t^\prime}\,dt^{\prime}dt^{\prime\prime}\,M_{a\gamma}^{\prime +}(t^{\prime})M_{a\gamma}^\times(t^{\prime\prime}) \right. \\ \left.\times e^{-i\left(\Delta M_1^\prime(t^\prime)-\Delta M_2(t^{\prime\prime})\right)} \right]\,e^{-i\tilde M_+^\prime(t)}\,A_\times(0)  + \left[1- \int_0^{t}\int_0^{t^\prime}\,dt^{\prime}dt^{\prime\prime}\,M_{a\gamma}^{\prime +}(t^{\prime})M_{a\gamma}^{\prime +}(t^{\prime\prime}) e^{-i\left(\Delta M_1^\prime(t^\prime)-\Delta M_1^\prime(t^{\prime\prime})\right)} \right. \\ \left.  - \int_0^{t}\int_0^{t^\prime}\,dt^{\prime}dt^{\prime\prime}\,M_F^\prime(t^{\prime})M_F^\prime(t^{\prime\prime}) e^{-i\left(\Delta M^\prime(t^\prime)-\Delta M^\prime(t^{\prime\prime})\right)} \right]e^{-i\tilde M_+^\prime(t)}A_+^\prime(0) \\   
+ \left[  \int_0^t dt^\prime\,M_F(t^\prime) e^{-i\Delta M^\prime(t^\prime)} - \int_0^{t}\int_0^{t^\prime}\,dt^{\prime}dt^{\prime\prime}\,M_{a\gamma}^{\prime +}(t^{\prime})M_{a\gamma}^{\times}(t^{\prime\prime}) e^{-i\left(\Delta M_1^\prime(t^\prime)-\Delta M_2^\prime(t^{\prime\prime})\right)} \right]e^{-i\tilde M_+^\prime(t)}A_\times^\prime(0) \\  + \left[  \int_0^t dt^\prime\,M_{a\gamma}^{\prime +}(t^\prime) e^{-i\Delta M_1^\prime(t^\prime)} + \int_0^{t}\int_0^{t^\prime}\,dt^{\prime}dt^{\prime\prime}\,M_F^\prime(t^{\prime})M_{a\gamma}^{\prime\times}(t^{\prime\prime}) e^{-i\left(\Delta M^\prime(t^\prime)+\Delta M_2^\prime(t^{\prime\prime})\right)} \right]e^{-i\tilde M_+^\prime(t)}a(0),
\end{gathered}
\end{equation}
\begin{equation}\nonumber
\begin{gathered}
A_\times^\prime(t) =- \left[\int_0^{t}\int_0^{t^\prime}\,dt^{\prime}dt^{\prime\prime}\,M_{a\gamma}^{\prime\times}(t^{\prime})M_{a\gamma}^{+}(t^{\prime\prime}) e^{-i\left(\Delta M_2^\prime(t^\prime)-\Delta M_1(t^{\prime\prime})\right)} \right]e^{-i\tilde M_\times^\prime(t)}A_+(0)-\left[ \int_0^{t}\int_0^{t^\prime}\,dt^{\prime}dt^{\prime\prime}\,M_{a\gamma}^{\prime\times}(t^{\prime})M_{a\gamma}^\times(t^{\prime\prime}) \right. \\ \left.\times e^{-i\left(\Delta M_2^\prime(t^\prime)-\Delta M_2(t^{\prime\prime})\right)} \right]\,e^{-i\tilde M_\times^\prime(t)}\,A_\times(0)  - \left[ \int_0^t dt^\prime\,M_F^\prime(t^\prime) e^{i\Delta M^\prime(t^\prime)} + \int_0^{t}\int_0^{t^\prime}\,dt^{\prime}dt^{\prime\prime}\,M_{a\gamma}^{\prime \times}(t^{\prime})M_{a\gamma}^{\prime +}(t^{\prime\prime})\right. \\ \left. \times  e^{-i\left(\Delta M_2^\prime(t^\prime)-\Delta M_1^\prime(t^{\prime\prime})\right)}\right]e^{-i\tilde M_\times^\prime(t)}A_+^\prime(0)   
+ \left[ 1 - \int_0^{t}\int_0^{t^\prime}\,dt^{\prime}dt^{\prime\prime}\,M_{a\gamma}^{\prime\times}(t^{\prime})M_{a\gamma}^{\prime \times}(t^{\prime\prime}) e^{-i\left(\Delta M_2^\prime(t^\prime)-\Delta M_2^\prime(t^{\prime\prime})\right)}  \right. \\ \left. - \int_0^{t}\int_0^{t^\prime}\,dt^{\prime}dt^{\prime\prime}\,M_F^\prime(t^{\prime})M_F^\prime(t^{\prime\prime}) e^{i\left(\Delta M^\prime(t^\prime)-\Delta M^\prime(t^{\prime\prime})\right)} \right]e^{-i\tilde M_\times^\prime(t)}A_\times^\prime(0)    \\
+  \left[  \int_0^t dt^\prime\,M_{a\gamma}^{\prime\times}(t^\prime) e^{-i\Delta M_2^\prime(t^\prime)} - \int_0^{t}\int_0^{t^\prime}\,dt^{\prime}dt^{\prime\prime}\,M_F^\prime(t^{\prime})M_{a\gamma}^{\prime +}(t^{\prime\prime}) e^{i\left(\Delta M^\prime(t^\prime)-\Delta M_1^\prime(t^{\prime\prime})\right)} \right]e^{-i\tilde M_\times^\prime(t)}a(0),
\end{gathered}
\end{equation}
\begin{equation}\label{time-dep-sol}
\begin{gathered}
a(t) =- \left[ \int_0^t dt^\prime\,M_{a\gamma}^+(t^\prime) e^{i\Delta M_1(t^\prime)} - \int_0^{t}\int_0^{t^\prime}\,dt^{\prime}dt^{\prime\prime}\,M_{a\gamma}^{\times}(t^{\prime})M_F(t^{\prime\prime}) e^{i\left(\Delta M_2(t^\prime)+\Delta M(t^{\prime\prime})\right)} \right]e^{-i\tilde M_a(t)}A_+(0)\\
- \left[ \int_0^t dt^\prime\,M_{a\gamma}^\times(t^\prime) e^{i\Delta M_2(t^\prime)} + \int_0^{t}\int_0^{t^\prime}\,dt^{\prime}dt^{\prime\prime}\,M_{a\gamma}^{+}(t^{\prime})M_F(t^{\prime\prime}) e^{i\left(\Delta M_1(t^\prime)-\Delta M(t^{\prime\prime})\right)} \right]e^{-i\tilde M_a(t)}A_\times(0)\\
- \left[ \int_0^t dt^\prime\,M_{a\gamma}^{\prime +}(t^\prime) e^{i\Delta M_1^\prime(t^\prime)} - \int_0^{t}\int_0^{t^\prime}\,dt^{\prime}dt^{\prime\prime}\,M_{a\gamma}^{\prime \times}(t^{\prime})M_F^\prime(t^{\prime\prime}) e^{i\left(\Delta M_2^\prime(t^\prime)+\Delta M^\prime(t^{\prime\prime})\right)} \right]e^{-i\tilde M_a(t)}A_+^\prime(0)\\
- \left[ \int_0^t dt^\prime\,M_{a\gamma}^{\prime \times}(t^\prime) e^{i\Delta M_2^\prime(t^\prime)} + \int_0^{t}\int_0^{t^\prime}\,dt^{\prime}dt^{\prime\prime}\,M_{a\gamma}^{\prime +}(t^{\prime})M_F^\prime(t^{\prime\prime}) e^{i\left(\Delta M_1^\prime(t^\prime)-\Delta M^\prime(t^{\prime\prime})\right)} \right]e^{-i\tilde M_a(t)}A_\times^\prime(0)\\
  + \left[ 1- \int_0^{t}\int_0^{t^\prime}\,dt^{\prime}dt^{\prime\prime}\,M_{a\gamma}^+(t^{\prime})M_{a\gamma}^{+}(t^{\prime\prime}) e^{i\left(\Delta M_1(t^\prime)-\Delta M_1(t^{\prime\prime})\right)} - \int_0^{t}\int_0^{t^\prime}\,dt^{\prime}dt^{\prime\prime}\,M_{a\gamma}^\times(t^{\prime})M_{a\gamma}^{\times}(t^{\prime\prime}) e^{i\left(\Delta M_2(t^\prime)-\Delta M_2(t^{\prime\prime})\right)}\right. \\ 
\left. - \int_0^{t}\int_0^{t^\prime}\,dt^{\prime}dt^{\prime\prime}\,M_{a\gamma}^{\prime +}(t^{\prime})M_{a\gamma}^{\prime +}(t^{\prime\prime}) e^{i\left(\Delta M_1^\prime(t^\prime)-\Delta M_1^\prime(t^{\prime\prime})\right)}  - \int_0^{t}\int_0^{t^\prime}\,dt^{\prime}dt^{\prime\prime}\,M_{a\gamma}^{\prime\times}(t^{\prime})M_{a\gamma}^{\prime\times}(t^{\prime\prime}) e^{i\left(\Delta M_2^\prime(t^\prime)-\Delta M_2^\prime(t^{\prime\prime})\right)}  \right]\times \\ e^{-i\tilde M_a(t)}a(0),
\end{gathered}
\end{equation}
where we have defined $\tilde M_\lambda(t)=\int dt \left(k(t)-M_\lambda(t)\right)$,  $\tilde M_\lambda^\prime(t)=\int dt \left(k(t)-M_\lambda^\prime(t)\right)$, $\tilde M_a(t)=\int dt \left(k(t)-M_a(t)\right)$ with $\lambda=(+, \times)$ and $\Delta M(t)= M_+ (t)- M_\times(t)$, $\Delta M_1(t)= M_+ (t)- M_a(t)$, $\Delta M_2(t)= M_\times (t)- M_a(t)$, $\Delta M^\prime(t)= M_+^\prime (t)- M_a(t)$,
 $\Delta M_2^\prime(t)= M_\times^\prime (t)- M_a(t)$.

Consider the amplitudes $A_\times(t)$ and $A_+(t)$ of ordinary photons as given in expression \eqref{time-dep-sol} and let us consider the case when $a(0)=0$. Therefore, we can write
\begin{align}\label{amplitudes}
A_+(t) &= I_1(t)A_+(0)-I_2(t)A_\times(0)-I_3(t)A_+^\prime(0)-I_4(t)A_\times^\prime(0)\nonumber \\
A_\times(t) &= -I_5(t)A_+(0)+I_6(t)A_\times(0)-I_7(t)A_+^\prime(0)-I_8(t)A_\times^\prime(0),
\end{align}
where we have defined
\begin{equation}\nonumber
\begin{gathered}
I_1(t) \equiv \left[1- \int_0^{t}\int_0^{t^\prime}\,dt^{\prime}dt^{\prime\prime}\,M_{a\gamma}^+(t^{\prime})M_{a\gamma}^+(t^{\prime\prime}) e^{-i\left(\Delta M_1(t^\prime)-\Delta M_1(t^{\prime\prime})\right)}   - \int_0^{t}\int_0^{t^\prime}\,dt^{\prime}dt^{\prime\prime}\,M_F(t^{\prime})M_F(t^{\prime\prime})\times\right.\\ \left.  e^{-i\left(\Delta M(t^\prime)-\Delta M(t^{\prime\prime})\right)} \right]e^{-i\tilde M_+(t)},\\
I_2(t)\equiv \left[  \int_0^t dt^\prime\,M_F(t^\prime) e^{-i\Delta M(t^\prime)} - \int_0^{t}\int_0^{t^\prime}\,dt^{\prime}dt^{\prime\prime}\,M_{a\gamma}^+(t^{\prime})M_{a\gamma}^\times(t^{\prime\prime}) e^{-i\left(\Delta M_1(t^\prime)-\Delta M_2(t^{\prime\prime})\right)} \right]\,e^{-i\tilde M_+(t)},\\ 
I_3(t)\equiv \left[ \int_0^{t}\int_0^{t^\prime}\,dt^{\prime}dt^{\prime\prime}\,M_{a\gamma}^+(t^{\prime})M_{a\gamma}^{\prime +}(t^{\prime\prime}) e^{-i\left(\Delta M_1^\prime(t^\prime)-\Delta M_1(t^{\prime\prime})\right)} \right]e^{-i\tilde M_+(t)},\\   
I_4(t)\equiv \left[ \int_0^{t}\int_0^{t^\prime}\,dt^{\prime}dt^{\prime\prime}\,M_{a\gamma}^+(t^{\prime})M_{a\gamma}^{\prime\times}(t^{\prime\prime}) e^{-i\left(\Delta M_1(t^\prime)-\Delta M_2^\prime(t^{\prime\prime})\right)} \right]e^{-i\tilde M_+(t)},\\
I_5(t)\equiv - \left[  \int_0^t dt^\prime\,M_F(t^\prime) e^{i\Delta M(t^\prime)} + \int_0^{t}\int_0^{t^\prime}\,dt^{\prime}dt^{\prime\prime}\,M_{a\gamma}^\times(t^{\prime})M_{a\gamma}^+(t^{\prime\prime}) e^{-i\left(\Delta M_2(t^\prime)-\Delta M_1(t^{\prime\prime})\right)} \right]\,e^{-i\tilde M_\times(t)},
\end{gathered}
\end{equation}

\begin{equation}\label{integrals}
\begin{gathered}
I_6(t)\equiv \left[1- \int_0^{t}\int_0^{t^\prime}\,dt^{\prime}dt^{\prime\prime}\,M_{a\gamma}^\times(t^{\prime})M_{a\gamma}^\times(t^{\prime\prime}) e^{-i\left(\Delta M_2(t^\prime)-\Delta M_2(t^{\prime\prime})\right)}   - \int_0^{t}\int_0^{t^\prime}\,dt^{\prime}dt^{\prime\prime}\,M_F(t^{\prime})M_F(t^{\prime\prime})\times\right.\\ \left.  e^{i\left(\Delta M(t^\prime)-\Delta M(t^{\prime\prime})\right)} \right]e^{-i\tilde M_\times(t)},\\
I_7(t)\equiv \left[ \int_0^{t}\int_0^{t^\prime}\,dt^{\prime}dt^{\prime\prime}\,M_{a\gamma}^\times(t^{\prime})M_{a\gamma}^{\prime +}(t^{\prime\prime}) e^{-i\left(\Delta M_2(t^\prime)-\Delta M_1^\prime(t^{\prime\prime})\right)} \right]e^{-i\tilde M_\times(t)},\\   
I_8(t)\equiv \left[ \int_0^{t}\int_0^{t^\prime}\,dt^{\prime}dt^{\prime\prime}\,M_{a\gamma}^\times(t^{\prime})M_{a\gamma}^{\prime\times}(t^{\prime\prime}) e^{-i\left(\Delta M_2(t^\prime)-\Delta M_2^\prime(t^{\prime\prime})\right)} \right]e^{-i\tilde M_\times(t)}.
\end{gathered}
\end{equation}

The intensity of ordinary photons in terms of the amplitudes $A_+(t)$ and $A_\times(t)$ is given by $I_\gamma(t)\equiv |A_+(t)|^2+|A_\times(t)|^2$. Using the expressions \eqref{amplitudes}, the absolute values of amplitudes square of the states $A_+(t)$ and $A_\times(t)$ are given by
\begin{equation}\nonumber
\begin{gathered}
|A_+(t)|^2=|I_1(t)|^2\,|A_+(0)|^2-2\,\text{Re}\left\{I_1(t)I_2^*(t)A_+(0)A_\times^*(0)\right\}-2\,\text{Re}\left\{I_1(t)I_3^*(t)A_+(0)A_+^{\prime *}(0)\right\} -\\ 2\,\text{Re}\left\{I_1(t)I_4^*(t)A_+(0)A_\times^{\prime *}(0)\right\} + |I_2(t)|^2\,|A_\times(0)|^2+2\,\text{Re}\left\{I_2(t)I_3^*(t)A_\times(0)A_+^{\prime *}(0)\right\} +2\,\text{Re}\left\{I_2(t)I_4^*(t)A_\times(0)A_\times^{\prime *}(0)\right\}\\
+ |I_3(t)|^2\,|A_+^\prime(0)|^2+2\,\text{Re}\left\{I_3(t)I_4^*(t)A_+^\prime(0)A_\times^{\prime *}(0)\right\} + |I_4(t)|^2\,|A_\times^\prime(0)|^2,\\
|A_\times(t)|^2=|I_5(t)|^2\,|A_+(0)|^2-2\,\text{Re}\left\{I_5(t)I_6^*(t)A_+(0)A_\times^*(0)\right\}+2\,\text{Re}\left\{I_5(t)I_7^*(t)A_+(0)A_+^{\prime *}(0)\right\} +\\ 2\,\text{Re}\left\{I_5(t)I_8^*(t)A_+(0)A_\times^{\prime *}(0)\right\} + |I_6(t)|^2\,|A_\times(0)|^2-2\,\text{Re}\left\{I_6(t)I_7^*(t)A_\times(0)A_+^{\prime *}(0)\right\} - 2\,\text{Re}\left\{I_6(t)I_8^*(t)A_\times(0)A_\times^{\prime *}(0)\right\}\\
+ |I_7(t)|^2\,|A_+^\prime(0)|^2+2\,\text{Re}\left\{I_7(t)I_8^*(t)A_+^\prime(0)A_\times^{\prime *}(0)\right\} + |I_8(t)|^2\,|A_\times^\prime(0)|^2.
\end{gathered}
\end{equation}

 \section{Photon polarization tensor in ordinary magnetized medium}
 \label{appendix-B}
 
When photons propagate in media, is well known that absorption and dispersive phenomena occur depending on the photon energy or frequency. In vacuum, the dispersion relation of photons is usually give by $\omega=\bs k^2$ where $\omega$ is the photon energy and $\bs k$ is the photon wave vector. However, in media such relation is modified in order to take into account the coherent interaction of photons with the medium. In general, in the presence of a medium, the vacuum Maxwell equations, $\Box A^\nu-\partial_\mu\partial^\nu A^\mu=0$, get modified to $\Box A^\nu-\partial_\mu\partial^\nu A^\mu=J^\nu$, in order to take into account the effects of the medium on the photons or electromagnetic waves. Here the current $J^\nu$ is the sum of external prescribed currents $J_\text{ext}^\nu$ and of the medium induced current $J_\text{ind}^\nu$.

Typically, if the fields propagating through the medium are sufficiently weak, one assumes a linear response of the medium due to the interaction of electromagnetic waves with external currents. In this case the induced current in momentum space can be written as power series of the four-potential $A^{\mu}(K)$
 \begin{equation}\label{lin-resp}
J_\text{ind}^\mu(K)=\Pi^{\mu\nu}(K)A_\nu(K) + \text{higher order terms},
\end{equation}
where $K=(\omega, \bs k)$ is the photon four-vector. The higher order terms reflect the non linear response of the medium. The first term in \eqref{lin-resp} is the one which defines the linear response of the medium where $\Pi^{\mu\nu}$ is the photon polarization tensor. 

The linear response term in \eqref{lin-resp} is the Fourier transform of a position space term \cite{Melrose} 
\begin{equation}\nonumber
J^\mu(x)=\int d^4 x^\prime\, \Pi^{\mu\nu}(x- x^\prime)\,A_\nu(x^\prime),
\end{equation}
where $\Pi^{\mu\nu}(x, x^\prime)=\Pi^{\mu\nu}(x- x^\prime)$ is the photon polarization tensor in position space of a homogeneous medium. The non locality of $\Pi^{\mu\nu}(x-x^\prime)$ follows from the fact that the relationships between the incident fields and the external currents are in general non local (within restrictions imposed by causality), see chap. 6 of Ref.  \cite{Raffelt:1996wa}. By keeping only the linear term in \eqref{lin-resp}, the modified Maxwell equations in momentum space in the presence of the medium become
\begin{equation}\label{mod-max}
(-\eta^{\mu\nu} K^2+K^\mu K^\nu+\Pi^{\mu\nu} )A_\nu=J_\text{ext}^\nu.
\end{equation}
It is straightforward to check that the modified Maxwell equations \eqref{mod-max} can be obtained by adding to the free electromagnetic Lagrangian density, the medium induced potential energy (or photon self energy in medium) in momentum space of the form $V=(1/2) A_\mu(K) \Pi^{\mu\nu}(K) A_\nu(K)$ or equivalently the Lagrangian density in position space of the form
\begin{equation}\label{lnd-curr}
\mathcal L_\text{med}(x)=-\frac{1}{2} \int d^4 x^\prime\, A_\mu (x) \Pi^{\mu\nu}(x-x^\prime)\,A_\nu(x^\prime).
\end{equation}

The explicit expression of the photon polarization tensor, that usually is calculated in momentum space, implicitly depends on the prescribed external currents that enter a given problem. In the case of a magnetized  medium, both classical and field theory expressions do exist. In this section, we focus on the calculation of the photon polarization tensor in a cold magnetized plasma. This kind of situation is quite common in many astrophysical and cosmological situations. Here, we derive gauge independent expressions for the polarization tensor by simply using classical arguments. The magnetized plasma is assumed to be with almost no collisions, globally neutral, anisotropic and homogeneous. There is no an external electric field. In the literature such an approximation is known as the Appleton approximation of the Drude model with a magnetic field added. The presence of the external magnetic field breaks the isotropy of the medium. 

Consider an electromagnetic wave which propagates along the $z$ axis in given cartesian coordinate system coincident with the coordinate system in which the plasma is at rest. Let the external magnetic field be $\bs B_e=B_e \hat{ \bs n}$ where $\hat{\bs n}=[\cos(\Theta), \sin(\Theta)\cos(\Phi), \sin(\Theta)\sin(\Phi)]$ as described in Sect. \ref{sec:2}. The induced motion on the $i$th electron in the plasma due to the combined action of the external magnetic field and incident electromagnetic wave is presumed to satisfy the classical equation of motion
\begin{equation}\label{el-eq}
m_e \ddot{\bs r}_i=-e \bs E-e\, \dot{\bs r}_i\times \bs B_e,
\end{equation}
where $\bs r_i$ is the position vector of the $i$th electron and $\bs E$ is the electric field of the incident electromagnetic wave. The contribution of the incident magnetic field wave is assumed to be negligible with respect to the prescribed external magnetic field $\bs B_e$. Now is more convenient to write the equation of motion \eqref{el-eq} in terms of the medium polarization vector $\bs P=-(e/V)\sum_i \bs r_i$ as
\begin{equation}\label{pol-eq}
\ddot{\bs P}= \omega_\text{pl}^2 \bs E-\omega_c\,\dot{\bs P}\times \hat{\bs n},
\end{equation}
where $\omega_\text{pl}=4 \pi \alpha n_e/m_e$ is the plasma frequency, $n_e$ is the free electron number density, $\omega_c=e B_e/m_e$ is the cyclotron frequency and $V$ is the volume of the region of space where the plasma is located. 
 
Assume that the fields evolve in time harmonically at a given point $\bs x$ and then let us write
\begin{equation}\label{field-exp-1}
\bs P(t)=\bs P(\omega) e^{-i\omega t}, \qquad \bs E(t)=\bs E(\omega) e^{-i\omega t}. 
\end{equation}
Now by inserting the expressions in \eqref{field-exp-1} into Eq. \eqref{pol-eq} and then solving for the components of $\bs P$, after lengthy calculations we get the following solution in terms of electric field components $E_j$
\begin{equation}
P_i=\chi_{ij} E_j, \qquad i, j=1, 2, 3.
\end{equation}
where $\chi_{ij}$ is the electromagnetic susceptibility tensor and the sum over repeated indexes is implicitly assumed. Its components in our case are given by
\begin{equation}
\begin{gathered}
\chi_{11}=-\frac{\omega_\text{pl}^2}{\omega^2-\omega_c^2}+\frac{\omega_\text{pl}^2\omega_c^2 \cos^2(\Theta)}{\omega^2(\omega^2-\omega_c^2)}, \quad \chi_{12}=\frac{\omega_\text{pl}^2\,\omega_c^2 \sin(2 \Theta)\cos(\Phi)}{2\, \omega^2(\omega^2-\omega_c^2)}+i \frac{\omega_\text{pl}^2\omega_c \sin(\Theta)\sin(\Phi)}{\omega(\omega^2-\omega_c^2)}, \\
\chi_{13}=\frac{\omega_\text{pl}^2\,\omega_c^2 \sin(2 \Theta)\sin(\Phi)}{2\,\omega^2(\omega^2-\omega_c^2)}-i \frac{\omega_\text{pl}^2\omega_c \sin(\Theta)\cos(\Phi)}{\omega(\omega^2-\omega_c^2)}, \quad \chi_{21} = \chi_{12}^*,\\
\chi_{22}=-\frac{\omega_\text{pl}^2}{\omega^2-\omega_c^2}+\frac{\omega_\text{pl}^2\omega_c^2 \sin^2(\Theta)\cos^2(\Phi)}{\omega^2(\omega^2-\omega_c^2)}, \quad \chi_{23} = \frac{\omega_\text{pl}^2\,\omega_c^2 \sin(2\Phi)\sin^2(\Theta)}{2\,\omega^2(\omega^2-\omega_c^2)} + i \frac{\omega_\text{pl}^2\omega_c \cos(\Theta)}{\omega(\omega^2-\omega_c^2)},\\
\chi_{31}=\chi_{13}^*, \quad \chi_{32}=\chi_{23}^*, \quad \chi_{33}=-\frac{\omega_\text{pl}^2}{\omega^2-\omega_c^2}+\frac{\omega_\text{pl}^2\omega_c^2 \sin^2(\Theta)\sin^2(\Phi)}{\omega^2(\omega^2-\omega_c^2)}.
\end{gathered}
\end{equation}

After having calculated the elements of electromagnetic susceptibility tensor, is quite straightforward to calculate the elements of the photon polarization tensor $\Pi^{ij}$. Indeed, their expressions are given by $\Pi^{ij}=-\chi^{ij}\, \omega^2$ for $(i, j=1, 2)$ and $\Pi^{ij}=-\chi^{ij}\, (\omega^2-k^2)^{1/2} \omega$ for $(ij=13, 23, 31, 32)$, and $\Pi^{33}=-\chi^{33}\, (\omega^2-k^2)$, see Refs. \cite{Melrose} and \cite{DOlivo:2002omk} for details. The matrix elements $\Pi^{11}$ and $\Pi^{22}$ correspond to the modification of the dispersion relations for the states $A_\times$ and $A_+$, namely the momentum space Maxwell equations become $\omega^2-k_{\times, +}^2=\omega^2(1-n_{\times, +}^2)=\Pi_{\times, +}$, where $n_{\times, +}$ are the total indexes of refraction and $\Pi^{11}=\Pi_\times, \Pi^{22}=\Pi_+$. The expressions for the elements $\Pi^{11}$ and $\Pi^{22}$ are given by
\begin{equation}\label{pol-ele}
\Pi^{11}=\frac{\omega^2 \omega_\text{pl}^2}{\omega^2-\omega_c^2} - \frac{\omega_\text{pl}^2\omega_c^2 \cos^2(\Theta)}{\omega^2-\omega_c^2}, \quad \Pi^{22}=\frac{\omega^2\omega_\text{pl}^2}{\omega^2-\omega_c^2} - \frac{\omega_\text{pl}^2\omega_c^2 \sin^2(\Theta)\cos^2(\Phi)}{\omega^2-\omega_c^2}.
\end{equation}

The first two terms in \eqref{pol-ele} correspond to the effect of only plasma to the polarization tensor. Indeed, it is straightforward to see that index of refraction corresponding to plasma frequency only (without external magnetic field) is given by $1- n_{\times, +}^2=\omega_\text{pl}^2/(\omega^2 - \omega_c^2)$. The second terms in \eqref{pol-ele} correspond to the Cotton-Mouton effect in plasma since this effect is proportional to $B_e^2$. The element $\Pi^{12}$ is given by
\begin{equation}
\Pi^{12}=-\frac{\omega_\text{pl}^2\,\omega_c^2 \sin(2 \Theta)\cos(\Phi)}{2\,(\omega^2-\omega_c^2)} - i \frac{\omega_\text{pl}^2\,\omega\,\omega_c \sin(\Theta)\sin(\Phi)}{\omega^2-\omega_c^2}.
\end{equation}
Since the element $\Pi^{12}$ is in general a complex quantity, it essentially means that the intensity of the state $A_\times$ changes for an electromagnetic wave propagating in magnetized plasma. The first term is due to the Cotton-Mouton effect while the second term corresponds to the Faraday effect in plasma. Typically in the literature one gets rid of the first term in $\Pi^{12}$ by choosing $\Phi=\pi/2$, namely by choosing the external magnetic field $\bs B_e$ and the photon wave-vector $\bs k$ in the $xz$ plane. In this case $\Pi^{12}$ is purely imaginary and it includes the Faraday effect only. 

The gauge invariant calculations for the elements of the polarization tensor presented above include also the longitudinal plasma oscillation, which in the Coulomb gauge modifies the dispersion relation for the scalar potential $A^0$. Indeed, in the Coulomb gauge the matrix element $\Pi^{33}$ is associated with the effective mass of the longitudinal plasma oscillation \cite{Raffelt:1996wa}. The presence of the longitudinal plasma oscillation makes the mixing of $A^0$ with the usual transverse photon states possible even in the case when there is no axion field. However, such mixing is usually very small since the elements of the photon polarization tensor for $(ij=13, 31, 23, 32)$ are proportional to $(\omega^2-k^2)^{1/2}$, which in general is a small quantity for $\omega\simeq k$. It is also worth to note that the elements of the photon polarization tensor in a cold magnetized plasma, calculated above by using a classical approach, do exactly coincide with the quantum field calculation of the photon polarization tensor in a cold magnetized plasma \cite{DOlivo:2002omk}.

\vspace{+2cm}

\end{document}